# Artificial intelligence pioneers the double-strangeness factory

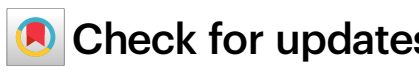

Yan He [1,2] ✉, Takehiko R. Saito [1,2,3,4] ✉, Hiroyuki Ekawa [2], Ayumi Kasagi [2,5], Yiming Gao [2,6,7], Enqiang Liu [2,6,7], Kazuma Nakazawa [2,8,9], Christophe Rappold [10], Masato Taki [5], Yoshiki K. Tanaka [2], He Wang [2,7,11], Ayari Yanai [2,3], Junya Yoshida [12] & Hongfei Zhang [1,13]

Artificial intelligence (AI) is transforming not only our daily experiences but also the technological development landscape and scientific research. In this study, we pioneered the application of AI in double-strangeness hypernuclear studies. Studies that investigate quantum systems with strangeness via hyperon interactions provide insights into fundamental baryon-baryon interactions and contribute to our understanding of the nuclear force and composition of neutron star cores. Specifically, we report the observation of a double-$\Lambda$ hypernucleus in nuclear emulsion achieved via innovative integration of machine learning techniques. The proposed methodology leverages generative AI and Monte Carlo simulations to produce training datasets combined with object detection AI for effective event identification. Based on the kinematic analysis and charge identification, the observed event was uniquely identified as the production and decay of ${}^{13}_{\Lambda\Lambda}$B, resulting from $\Xi^-$ capture by $^{14}$N in the nuclear emulsion. Assuming $\Xi^-$ capture in the atomic 3D state, the binding energy of the two $\Lambda$ hyperons in ${}^{13}_{\Lambda\Lambda}$B, $B_{\Lambda\Lambda}$, was determined as $25.57 \pm 1.18(\text{stat.}) \pm 0.07(\text{syst.})$ MeV. The $\Lambda\Lambda$ interaction energy $\Delta B_{\Lambda\Lambda}$ obtained was $2.83 \pm 1.18(\text{stat.}) \pm 0.14(\text{syst.})$ MeV.

The rapid advancements in artificial intelligence (AI) in recent years have transformed our daily lives. The integration of AI with established technologies is particularly notable[1]. In many cases, such fusion has revitalized long-standing methods and transformed them into cutting-edge tools. This synergy not only modernizes traditional technologies but also opens up entirely new avenues of exploration.

Such progress in AI is now beginning to contribute significantly to addressing some of humanity's most enduring questions: What is the nature of matter? How did the Universe begin? In the fields of quarks and nuclear physics, which have made substantial contributions to our understanding of both matter and the cosmos, the application of AI has yielded novel and promising insights. We sought to harness AI in combination with nuclear emulsion technology, a well-established and mature technique, to advance our understanding of quantum many-body systems involving two strange quarks with a strangeness degree of freedom, called the double-strangeness ($S = -2$) sector. This combination allows for detecting diverse double-strangeness hypernuclei that have long

[1]School of Nuclear Science and Technology, Lanzhou University, Lanzhou, Gansu, China. [2]High Energy Nuclear Physics Laboratory, RIKEN, Wako, Saitama, Japan. [3]Department of Physics, Saitama University, Saitama, Japan. [4]GSI Helmholtz Centre for Heavy Ion Research, Darmstadt, Germany. [5]Graduate School of Artificial Intelligence and Science, Rikkyo University, Tokyo, Japan. [6]Institute of Modern Physics, Chinese Academy of Sciences, Lanzhou, Gansu, China. [7]University of Chinese Academy of Sciences, Beijing, China. [8]Faculty of Education, Gifu University, Gifu, Japan. [9]The Research Institute of Nuclear Engineering, University of Fukui, Fukui, Japan. [10]Instituto de Estructura de la Materia CSIC, Madrid, Spain. [11]State Key Laboratory of Heavy Ion Science and Technology, Institute of Modern Physics, Chinese Academy of Sciences, Lanzhou, Gansu, China. [12]International Center for Synchrotron Radiation Innovation Smart, Tohoku University, Sendai, Japan. [13]School of Physics, Xi'an Jiaotong University, Xi'an, Shaanxi, China. ✉e-mail: yhe21@lzu.edu.cn; takehiko.saito@riken.jp

eluded observation and lays the foundation for constructing a "double-strangeness factory".

Double-strangeness quantum many-body systems are fundamental to nuclear, particle, and astrophysics, particularly in understanding neutron stars and ultra-dense supernova remnants with radii of ~10 km but masses approximately twice that of the Sun[2]. Despite advances in gravitational-wave detections and r-process studies[3,4], their internal structure remains elusive. At such densities, matter likely shifts from nucleons to strange-quark-containing hyperons, making hyperon–nucleon and hyperon–hyperon (YY) interactions crucial for the equation of state (EOS). However, limited experimental data, particularly for YY interactions, render the EOS too soft[5], which conflicts with observations of massive neutron stars. This mismatch, known as the "hyperon puzzle", remains unresolved. Addressing this issue requires precise data on YY interactions, which are scarce.

Quantum chromodynamics predicts an attractive short-range baryon–baryon interaction in the double-strange sector, in contrast to the repulsive core of the nucleon–nucleon interaction[6,7]. A prominent double-strangeness system, the H-dibaryon (a six-quark state: uuddss), remains the subject of intense research[8]. Despite extensive experimental research[9–12], definitive observations are lacking. Theoretical studies have suggested that H-dibaryon may exist within double-strangeness hypernuclei[13–15] and neutron star cores[16]. Although direct observation of nuclear matter is lacking, the measured masses of double-Λ hypernuclei provide crucial constraints on the mass of H-dibaryon[17–20]. Investigating the dependence of the ΛΛ interaction on the mass number of double-Λ hypernuclei can provide insights into the existence of H-dibaryon[21].

Double-strangeness hypernuclei, containing two strange quarks, which occupy the $S = -2$ sector and act as "small experimental laboratories" for probing hyperon interactions in nuclei, are a key focus in the context of $S = -2$ systems. In particular, a double-Λ hypernucleus with two Λ hyperons in the nucleus provides key insights into the ΛΛ interaction[22,23]. The ΛΛ interaction, confirmed to be weakly attractive through observations of double-Λ hypernuclei, is the only known YY interaction experimentally. In addition, the binding of the three-body ΛΛN system, particularly the ΛΛd configuration, remains unresolved, although several theoretical models predict a bound particle-stable ${}^{4}_{\Lambda\Lambda}$H if the ΛΛ interaction is sufficiently attractive[24–26]. However, experimental confirmation is lacking because the initial claim from the BNL-AGS E906 experiment[27] was later downgraded[28].

Despite over 70 years of hypernuclear research since it was first observed in 1953[29], only 47 double-strangeness hypernucleus candidates have been observed using nuclear emulsion. Notably, the NAGARA event[19,20] remains the only confirmed observation of ${}^{6}_{\Lambda\Lambda}$He, the lightest double-Λ hypernucleus. Its fully occupied s-shell reveals a weak (<1 MeV), attractive s-wave ΛΛ interaction, which serves as a critical benchmark for YY interaction calculations and interpretation of less definitive events. However, to date, the NAGARA event remains the only uniquely identified double-Λ hypernucleus, with other double-Λ hypernuclei exhibiting ambiguous identifications. Therefore, information on the ΛΛ interaction in the nuclear medium is not yet precisely known, and substantially more experimental data are required. The short hyperon lifetime (~$10^{-10}$s) hinders scattering experiments, particularly for multi-hyperon systems. Nuclear emulsion experiments with their sub-micrometer resolution[30] offer the unique advantage of visualizing decay chains, enabling nuclide identification through event-by-event analysis of double-Λ hypernuclear production and decays.

The J-PARC E07 experiment[21], a recent hybrid-emulsion study at the Japan Proton Accelerator Research Complex (J-PARC), aims to detect approximately $10^2$ double-Λ hypernuclei events to broaden our understanding of strangeness physics. The experiment used a 1.81 GeV/$c$ $K^-$ beam at the K1.8 beam line of the Hadron Experimental Facility at J-PARC. $\Xi^-$ hyperons were produced via the quasifree "$p$"($K^-$, $K^+$)$\Xi^-$ reaction on a diamond target and subsequently injected into an emulsion module positioned downstream of the target. In the nuclear emulsion, the $\Xi^-$ hyperons captured by nuclei within the emulsion stacks were subsequently tracked, while the associated $K^+$ were tracked using silicon strip detectors to measure their positions and angles. However, the detection efficiency for all double-strangeness hypernuclear events in emulsion sheets has been estimated to be approximately 10%[31,32], mainly due to the limited spectrometer acceptance and tracking, as well as the fact that the cross-section of the reaction "$n$"($K^-$, $K^0$)$\Xi^-$, which cannot be detected using the hybrid method, is about twice that of "$p$"($K^-$, $K^+$)$\Xi^-$[33]. Although 33 candidate events triggered by $\Xi^-$ hyperons were detected, only three events could be identified: Mino[34], Ibuki[35], and Irrawaddy[36]. Unfortunately, no double-Λ hypernucleus was uniquely identified. Nervertheless, the entire emulsion volume is estimated to contain over a thousand double-strangeness events[33], including untriggered events generated by the "$n$"($K^-$, $K^0$)$\Xi^-$ reaction. Therefore, it is essential to develop a new and efficient detection method for mining these events.

In this study, we developed an efficient analysis pipeline based on machine learning techniques for detecting double-Λ hypernuclear events, characterized by a "three-vertex" topology, within nuclear emulsion from the J-PARC E07 experiment[37]. Here, we report the first uniquely identified observation of a double-Λ hypernucleus ${}^{13}_{\Lambda\Lambda}$B achieved using our AI-driven nuclear emulsion approach; this is also the second unique identification in history, 24 years after the discovery of the NAGARA event[19]. This success paves the way for further discoveries of double-strangeness hypernuclei, ultimately establishing a "Double-Strangeness Factory".

## Results

We developed a machine learning method[37] that combines generative adversarial networks (GANs)[38] and Geant4 Monte Carlo simulations[39] for training data generation, along with a Mask Region-based Convolutional Neural Network (Mask R-CNN)[40] for object detection, as discussed in Section "Methods". Our model has already detected six double-Λ hypernuclear event candidates, one of which was uniquely identified, as discussed below; this is the second uniquely identified double-Λ hypernucleus since the discovery of hypernuclei with nuclear emulsion in 1953.

Figure 1 illustrates this event. Following $\Xi^-$ capture at vertex A, a double-Λ hypernucleus is produced, undergoing sequential cascade decays at vertices B and C. All the emitted particles were brought to rest within the emulsion. The range and emission angle measurements for each particle are detailed in Table 1. A detailed description is provided in Fig. 1 and Table 1.

We analyzed the single-Λ hypernuclear (track #2) decay at vertex C. Three charged particles (tracks #6, #7, and #8) are emitted from vertex C. The coplanarity, calculated as $(\vec{n_1} \times \vec{n_2}) \cdot \vec{n_3}$, where $\vec{n_i}$ represents the unit vector of the angles of tracks #6, #7, and #8, respectively, was measured as $0.004 \pm 0.013$. This value suggests that the three particles were emitted within a single plane, significantly reducing the likelihood of neutron emission. The constraints of the energy and momentum conservation are as follows:

$$M_{\#2} = \sum_{\#6,\#7,\#8} \sqrt{M_i^2 + |\vec{P}_i|^2}$$
$$\vec{P}_m + \vec{P}_{\#6} + \vec{P}_{\#7} + \vec{P}_{\#8} = \vec{0} \quad (1)$$

where $M_i$ and $\vec{P}_i$ denote the mass and momentum of the particle for tracks #6, #7 and #8, respectively, and $\vec{P}_m$ denotes the missing momentum. For momentum balance, we defined $\chi^2$ as:

$$\chi^2 = \left(\frac{P_{mx}}{\sigma_{mx}}\right)^2 + \left(\frac{P_{my}}{\sigma_{my}}\right)^2 + \left(\frac{P_{mz}}{\sigma_{mz}}\right)^2 \quad (2)$$

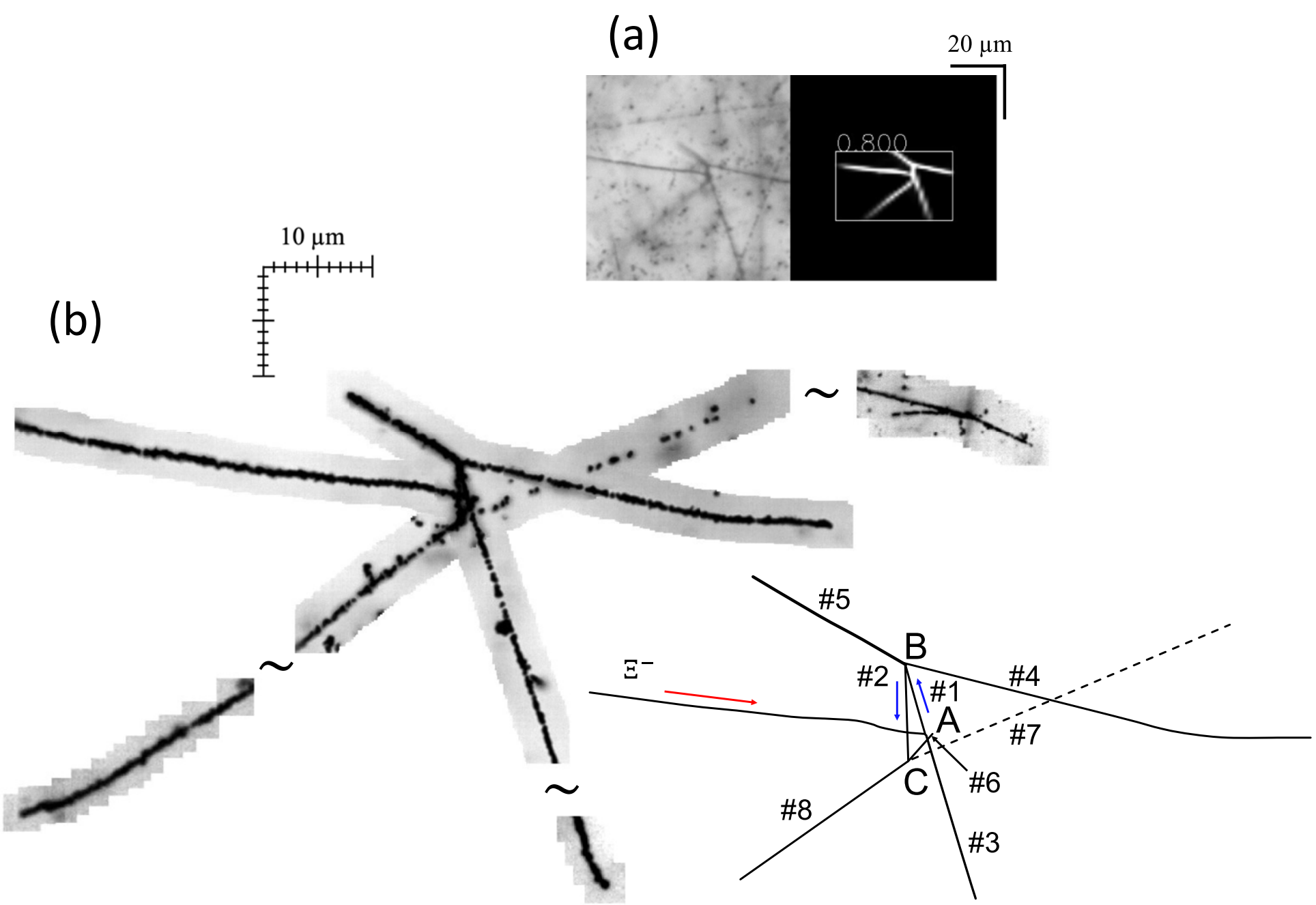

**Fig. 1 | Detection of the double-Λ hypernuclear event using machine learning.** **a** The detection result of the event from the machine learning analysis pipeline. The left panel depicts the input emulsion image, acquired under an optical microscope with a 20× objective lens, and the right panel presents the model's output, highlighting the event topology with a confidence score of 0.8. **b** Photograph and schematic diagram of the event. An incoming $\Xi^-$ particle (red arrow) is captured at vertex A, producing a double-Λ hypernucleus (#1) and particle #3. Cascade decay (blue arrows) of double-Λ hypernucleus (#1) occurred at vertices B and C, with emitted particles #4 and #5 from B, and #6, #7, and #8 from C, all stopping in the emulsion. Track #7 is identified as a $\pi^-$ particle, represented by a dashed line. Particle length and angle measurements are detailed in Table 1.

$P_{mx}$, $P_{my}$, and $P_{mz}$ are the three components of the missing momentum and $\sigma_{mx}$, $\sigma_{my}$ and $\sigma_{mz}$ are the corresponding errors. Momentum conservation was considered satisfied when $\chi^2$ was less than 13.898. This $\chi^2$ threshold, with a degree of freedom of 3, corresponds to a $p$-value of 0.003. In addition, after considering the energy conservation within the $3\sigma$ confidence level, all the decay modes at vertex C are listed in Table 2. The momentum and kinetic energy of each particle were obtained using the calibrated range-energy relationship described in Section "Methods". Given the observed coplanarity and thus the assumption of no neutron emission, ${}^{5}_{\Lambda}$He and ${}^{9}_{\Lambda}$He remain as possible decay modes. However, the termination of track #6 without any visible decay products, as depicted in Fig. 1b, rules out the possibility of it being $^{8}$He for the decay mode of ${}^{9}_{\Lambda}$He; this is because $^{8}$He undergoes a beta decay[41,42], which produces visible tracks of electron and lighter nuclei in nuclear emulsion.

**Table 1 | Measured ranges and emission angles of particles from the observed double-Λ hypernuclear event**

| Vertex | Track ID | Range [μm] | θ [°] | ϕ [°] | Comment |
|---|---|---|---|---|---|
| A | #1 | 3.5 ± 0.2 | 100.9 ± 2.3 | 279.7 ± 2.2 | double-Λ hypernucleus |
| | #3 | 757.1 ± 3.0 | 93.4 ± 1.4 | 109.1 ± 1.1 | |
| B | #2 | 7.0 ± 0.3 | 124.6 ± 1.1 | 89.6 ± 1.3 | single-Λ hypernucleus |
| | #4 | 36.8 ± 0.2 | 83.9 ± 1.7 | 168.9 ± 1.1 | |
| | #5 | 22.4 ± 0.4 | 145.5 ± 0.8 | 326.7 ± 1.1 | |
| C | #6 | 3.0 ± 0.4 | 52.3 ± 3.1 | 234.4 ± 3.8 | |
| | #7 | 7912.8 ± 3.0 | 54.0 ± 0.3 | 202.5 ± 0.3 | $\pi^-$ |
| | #8 | 1105.5 ± 3.0 | 127.2 ± 1.0 | 36.36 ± 0.9 | |

Angles are expressed by a zenith angle ($\theta$) with respect to the direction perpendicular to the plate and an azimuthal angle ($\phi$). The error arises from the measurement uncertainties due to image-based fitting. The event's topology is consistent with the sequential weak decay of a double-Λ hypernucleus, where track #1 is identified as a double-Λ hypernucleus and track #2 as a single-Λ hypernucleus. Track #7 is identified as a $\pi^-$ meson based on the observation of its capture star at its endpoint in the emulsion.

While the coplanarity of the three tracks at vertex C suggests planar emission for single-Λ hypernucleus decay, low-probability neutron emission within the plane remains possible. With the energy and momentum conservation as mentioned above, neutron emission from ${}^{3}_{\Lambda}$H, ${}^{5}_{\Lambda}$H, ${}^{5}_{\Lambda}$He, and ${}^{10}_{\Lambda}$He is kinematically allowed, yet experimental evidence of those decay branches is scarce. For decays involving neutron emission, the missing momentum was assigned to a neutron ($\overrightarrow{P}_m = \overrightarrow{P}_n$), where $\overrightarrow{P}_n$ represents its momentum. For ${}^{3}_{\Lambda}$H, fewer than 30 neutron-emission decays were observed out of 2000 events in previous emulsion experiments[43], which is consistent with the theoretical predictions of a 0.6% branching ratio[44]. ${}^{5}_{\Lambda}$H remains unobserved, which suggests a low probability of its existence as a bound state. Although ${}^{6}_{\Lambda}$He was detected[45], all 31 observed decays were mesonic without neutron emission, similar to ${}^{5}_{\Lambda}$He (1784 events), as depicted in Table 2. Finally, ${}^{10}_{\Lambda}$He decay, involving $^{8}$He as a daughter nucleus, was excluded because of the absence of the characteristic shape of the tracks associated with the decay of $^{8}$He. Given the value of coplanarity and previous experimental evidences, the single-Λ hypernucleus (#2) is most likely ${}^{5}_{\Lambda}$He, and its decay mode at vertex C is:

$$ {}^{5}_{\Lambda}\mathrm{He} \rightarrow {}^{4}\mathrm{He} + \pi^- + p \quad (3) $$

The production mode of the double-Λ hypernucleus at vertex A was analyzed. A $\Xi^-$ was captured at rest in the emulsion, emitting two charged particles (tracks #1 and #3). Although the emulsion contains both heavy (Ag, Br) and light (C, N, O) nuclei, the short range of track #1 (3.5 ± 0.2 μm) implies that its energy was insufficient to overcome the Coulomb barrier of the heavy nuclei[18]. Additionally, capture by heavy nuclei typically results in the emission of Auger electrons[46], which were

**Table 2 | Possible decay modes of the single-Λ hypernucleus (#2) at vertex C**

| #2 | | #6 | #7 | #8 | Neutron | $B_\Lambda$[MeV] | $B^p_\Lambda$[MeV] | Comment |
|---|---|---|---|---|---|---|---|---|
| $^{5}_{\Lambda}$He | → | $^{4}$He | $\pi^-$ | $p$ | | 2.96 ± 0.37 | 3.12 ± 0.02 | most probable |
| $^{9}_{\Lambda}$He | → | $^{8}$He | $\pi^-$ | $p$ | | 3.56 ± 0.37 | 6.60 ± 0.12 | rejected |
| $^{3}_{\Lambda}$H | → | $p$ | $\pi^-$ | $p$ | $n$ | −0.78 ± 0.42 | 0.13 ± 0.05 | |
| $^{5}_{\Lambda}$H | → | $t$ | $\pi^-$ | $p$ | $n$ | 3.99 ± 0.44 | 3.95 ± 0.50 | |
| $^{6}_{\Lambda}$He | → | $^{4}$He | $\pi^-$ | $p$ | $n$ | 3.97 ± 0.42 | 4.18 ± 0.10 | |
| $^{10}_{\Lambda}$He | → | $^{8}$He | $\pi^-$ | $p$ | $n$ | 4.87 ± 0.39 | 7.46 ± 0.15 | rejected |

Λ binding energies ($B_\Lambda$) are derived from mass reconstruction of the single-Λ hypernucleus at vertex C, with errors determined via kinematic fitting[62] using measurement information. $B^p_\Lambda$ represents binding energies from prior emulsion experiments[45]. The $B^p_\Lambda$ values for $^{9}_{\Lambda}$He, $^{5}_{\Lambda}$H, and $^{10}_{\Lambda}$H are estimated by linear interpolation of the isotopic trend, with uncertainties of ±1 MeV adopted in the analyses, which significantly exceed the prediction intervals.

**Table 3 | Possible production modes of double-Λ hypernucleus (#1) at vertex A**

| $\Xi^-$ capture | | #1 | #3 | Neutron | $\Delta B_{\Lambda\Lambda} - B_{\Xi^-}$ [MeV] | Comment |
|---|---|---|---|---|---|---|
| $\Xi^-$ + $^{12}$C | → | $^{10}_{\Lambda\Lambda}$Be | $p$ | 2$n$ | > 9.51 ± 0.36 | |
| $\Xi^-$ + $^{12}$C | → | $^{10}_{\Lambda\Lambda}$Be | $d$ | $n$ | 11.75 ± 0.77 | |
| $\Xi^-$ + $^{12}$C | → | $^{11}_{\Lambda\Lambda}$Be | $p$ | $n$ | 6.55 ± 0.71 | |
| $\Xi^-$ + $^{14}$N | → | $^{11}_{\Lambda\Lambda}$B | $p$ | 3$n$ | > 19.58 ± 0.38 | |
| $\Xi^-$ + $^{14}$N | → | $^{11}_{\Lambda\Lambda}$B | $d$ | 2$n$ | > 19.76 ± 0.50 | |
| $\Xi^-$ + $^{14}$N | → | $^{12}_{\Lambda\Lambda}$B | $p$ | 2$n$ | > 10.41 ± 0.57 | |
| $\Xi^-$ + $^{14}$N | → | $^{12}_{\Lambda\Lambda}$B | $d$ | $n$ | 10.48 ± 0.99 | |
| $\Xi^-$ + $^{14}$N | → | $^{13}_{\Lambda\Lambda}$B | $p$ | $n$ | 2.66 ± 1.18 | most probable |
| $\Xi^-$ + $^{16}$O | → | $^{14}_{\Lambda\Lambda}$C | $p$ | 2$n$ | > 8.50 ± 0.86 | |
| $\Xi^-$ + $^{16}$O | → | $^{14}_{\Lambda\Lambda}$C | $d$ | $n$ | 7.99 ± 1.38 | |
| $\Xi^-$ + $^{16}$O | → | $^{15}_{\Lambda\Lambda}$C | $p$ | $n$ | 13.80 ± 1.78 | |

Light nucleus capture (e.g. by $^{12}$C, $^{14}$N, $^{16}$O) in the emulsion layer is considered. Errors in $\Delta B_{\Lambda\Lambda} - B_{\Xi^-}$ are decided by the results of the kinematic fitting. Only candidates with $\Delta B_{\Lambda\Lambda} - B_{\Xi} < 20$ MeV are listed.

not observed in this event. Thus, our analysis considered capture by $^{12}$C, $^{14}$N, or $^{16}$O in double-Λ hypernucleus production.

From Table 1, tracks #1 and #3 are not back-to-back with $p$-value cut condition of 0.003, which indicates at least one neutron emission. After considering momentum and energy conservation as mentioned above, the possible production modes are listed in Table 3. For decays with neutron emission, the missing momentum was assigned to the neutron(s). In the cases of multiple neutron emissions, it was assumed that the neutrons share the same momentum, producing a minimum kinetic energy and a lower limit for $\Delta B_{\Lambda\Lambda} - B_{\Xi}$, where $\Delta B_{\Lambda\Lambda}$ is the ΛΛ interaction energy and $B_{\Xi}$ is the $\Xi^-$ binding energy. $\Delta B_{\Lambda\Lambda}$ is derived from the binding energies of the single-Λ ($B_\Lambda$) and double-Λ ($B_{\Lambda\Lambda}$) hypernuclei:

$$\Delta B_{\Lambda\Lambda}\left(^{A}_{\Lambda\Lambda}\mathrm{Z}\right) = B_{\Lambda\Lambda}\left(^{A}_{\Lambda\Lambda}\mathrm{Z}\right) - 2B_{\Lambda}\left(^{A-1}_{\Lambda}\mathrm{Z}\right) \quad (4)$$

Most of the production modes in Table 3 result in large $\Delta B_{\Lambda\Lambda} - B_{\Xi}$ values, except for $^{13}_{\Lambda\Lambda}$B. Notably, all measured $\Delta B_{\Lambda\Lambda}$ values, including uniquely identified NAGARA event with $\Delta B_{\Lambda\Lambda} = 0.67 \pm 0.17$[20] MeV, and other double-Λ hypernuclear events with ambiguous identifications, were below 5 MeV[47]. This is consistent with the femtoscopy results from ALICE[48], which estimated $\Delta B_{\Lambda\Lambda} = 3.2^{+1.6}_{-2.4}(\mathrm{stat})^{+1.8}_{-1.0}(\mathrm{syst})$ MeV for ΛΛ bound state. With this constraint of 5 MeV for ΛΛ interaction, only $^{13}_{\Lambda\Lambda}$B is most likely to produce a double-Λ hypernucleus at vertex A. However, the $\Delta B_{\Lambda\Lambda}$ values for $^{11}_{\Lambda\Lambda}$Be and $^{14}_{\Lambda\Lambda}$C could also be consistent with this constraint within three-sigma uncertainty.

Although the production analysis at vertex A suggests $^{13}_{\Lambda\Lambda}$B as the most likely double-Λ hypernucleus, we further analyzed its decay at vertex B. Track #2 was identified as $^{5}_{\Lambda}$He from vertex C. The emission of tracks #4 and #5 at vertex B indicates that their total charge should be 3$e$ and that one track must have a larger charge than the other if the analyses at vertices A and C are consistent.

To identify the charges of tracks #4 and #5 of the newly observed double-Λ hypernuclear event, as illustrated in Fig. 1, we employed the method described in Ref. 49. We collected 57 $\alpha$ tracks from the $\alpha$ decay chain, all with a $\theta$ angle similar to that of track #4 in three degrees and 50 $\alpha$ tracks associated with track #5. Fig. 2a, b show the distribution of the measured volume of these $\alpha$ tracks relative to tracks #4 (a) and #5 (b), respectively, at different distances from the stopping point to the initial point in the emulsion sheet. In nuclear emulsion, the track volume is correlated with the energy deposition. As tracks are formed by developed grains, particles with different charges exhibit variations in both track thickness and volume[49].

Figure 2a, b shows that track #4 exhibits smaller track volumes than the $\alpha$ tracks at various distances, whereas track #5 displays a volume distribution identical to that of the $\alpha$ tracks. From these volume distributions, a statistical hypothesis test was conducted to determine whether tracks #4 and #5 were similar to the $\alpha$ tracks (null hypothesis, H0) or not (alternative hypothesis, H1). The results of the statistical test are depicted in Fig. 2c, d as a function of the range of tracks #4 and track #5. These figures illustrate the confidence level for rejecting the H0 hypothesis (i.e., similarity to the $\alpha$ track) for tracks #4 and #5. Fig. 2c shows an increasing confidence level for rejecting H0 as the range of track #4 increases, ultimately reaching a cumulative confidence level of 2.4$\sigma$ at the maximum range of track #4. Conversely, Fig. 2d exhibits no significant confidence in rejecting H0 across the entire range of track #5. This evidence demonstrates that track #4 has a smaller charge than $\alpha$ nuclei, whereas track # 5 has the same charge as $\alpha$ nuclei.

Kinematic analysis at vertex B, shown in Table 4, lists possible decay modes for $^{11}_{\Lambda\Lambda}$Be, $^{13}_{\Lambda\Lambda}$B, and $^{14}_{\Lambda\Lambda}$C. The decay modes of $^{13}_{\Lambda\Lambda}$B are unique owning to the inclusion of tracks #4 (Z=1) and #5 (Z=2), which align with the charge identification from both the track volume analysis at vertex B and the analyses at vertices A and C.

Based on the analyses at vertices A, B, and C, the newly observed double-Λ hypernucleus is uniquely identified as $^{13}_{\Lambda\Lambda}$B, which is produced by the reaction described in Eq. (5).

$$\Xi^- + {}^{14}\mathrm{N} \rightarrow {}^{13}_{\Lambda\Lambda}\mathrm{B} + p + n \quad (5)$$

Because of the multi-neutron emission of the decay of the double-Λ hypernucleus, only the upper limit of the $\Delta B_{\Lambda\Lambda}$ value can be determined, as listed in Table 4. Therefore, $\Delta B_{\Lambda\Lambda} - B_{\Xi^-} = 2.66 \pm 1.18$ MeV, which is constrained by the production mode at vertex A.

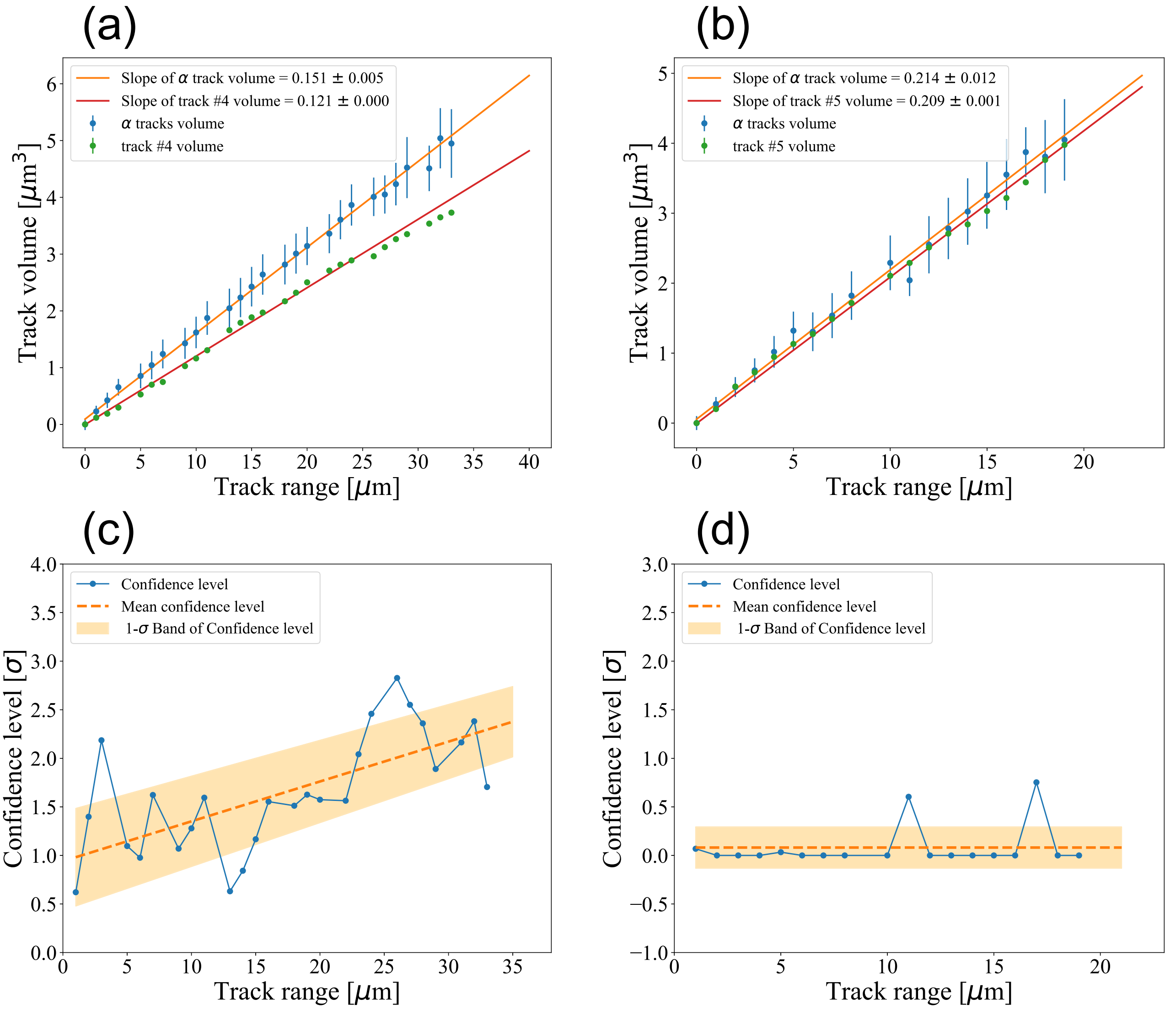

**Fig. 2 | Charge identification for tracks #4 and #5.** The distribution of $\alpha$ track volumes relative to that of track #4 (**a**) and track #5 (**b**), respectively, at different distances from the stopping point to the initial point of tracks in the emulsion sheet. From these volume distributions, a statistical hypothesis test was conducted to determine whether tracks #4 and #5 were similar to the alpha tracks (null hypothesis, H0) or not (alternative hypothesis, H1). **c** clearly displays an increasing confidence level for rejecting H0 as the range of track #4 increases, ultimately reaching a cumulative confidence level of 2.4$\sigma$ at the maximum range of track #4. In contrast, **d** exhibits no significant confidence in rejecting H0 across the entire range of track #5.

According to theoretical calculations for the nuclear absorption rate of $\Xi^-$ hyperons, $\Xi^-$ hyperon capture from an atomic 3D state is dominant[50,51], whereas only a small percentage of the probability was estimated for 2P state capture[52]. Thus, we assumed that the $\Xi^-$ capture in the atomic 3D state of $^{14}$N is most probable. Using a theoretical $B_{\Xi^-}$ value of 0.174 MeV[50,51], which is primarily determined by the Coulomb potential, we obtained the binding energy of two Λ hyperons in $^{13}_{\Lambda\Lambda}$B:

$$B_{\Lambda\Lambda} = 25.57 \pm 1.18(\text{stat.}) \pm 0.07(\text{syst.})\text{MeV} \tag{6}$$

and the ΛΛ interaction energy is:

$$\Delta B_{\Lambda\Lambda} = 2.83 \pm 1.18(\text{stat.}) \pm 0.14(\text{syst.})\text{MeV} \tag{7}$$

Statistical errors arise from kinematic fitting, while systematic errors are due to the masses of the $\Xi^-$ hyperon (1321.71 ± 0.07 MeV), Λ hyperon (1115.683 ± 0.006 MeV)[53], and $B_\Lambda(^{12}_{\Lambda}\text{B}) = 11.37 \pm 0.06$ MeV[45].

## Discussion

Using advanced machine learning, we analyzed 0.2% of the E07 emulsion data and achieved the first unambiguous identification of a double-Λ hypernucleus $^{13}_{\Lambda\Lambda}$B. Extrapolating from the observed candidate rate, the entire dataset is expected to contain more than 2000 double-strangeness hypernuclear events, with hundreds still awaiting identification. This work marks the dawn of a double-strangeness factory, where AI-driven analysis unlocks unprecedented access to rare double-strangeness hypernuclei.

The ΛΛ binding energy in $^{13}_{\Lambda\Lambda}$B was determined as $B_{\Lambda\Lambda} = 25.57 \pm 1.18(\text{stat.}) \pm 0.07(\text{syst.})$ MeV with an interaction energy of $\Delta B_{\Lambda\Lambda} = 2.83 \pm 1.18(\text{stat.}) \pm 0.14(\text{syst.})$ MeV. This result confirms the existence of $^{13}_{\Lambda\Lambda}$B and is the second unambiguous identification of a double-Λ hypernucleus since the first observation of hypernuclei in 1953[29].

Previous experiments have suggested possible interpretations of $^{13}_{\Lambda\Lambda}$B; however, without unique identification, definitive conclusions regarding its binding properties were prevented. The E176 experiment[47] reported $B_{\Lambda\Lambda} = 23.3 \pm 0.7$ MeV and $\Delta B_{\Lambda\Lambda} = 0.6 \pm 0.8$ MeV,

**Table 4 | Possible decay modes of the double-Λ hypernucleus based on kinematic analysis at vertex B**

| #1 → #2 + #4 + #5 | $\Delta B_{\Lambda\Lambda}$ [MeV] |
|---|---|
| ${}^{11}_{\Lambda\Lambda}$Be → ${}^{5}_{\Lambda}$He + $p$ + $d$ + 3$n$ | $<122.59 \pm 0.26$ |
| ${}^{11}_{\Lambda\Lambda}$Be → ${}^{5}_{\Lambda}$He + $p$ + $t$ + 2$n$ | $<124.47 \pm 0.39$ |
| ${}^{11}_{\Lambda\Lambda}$Be → ${}^{5}_{\Lambda}$He + $d$ + $p$ + 3$n$ | $< 123.11 \pm 0.28$ |
| ${}^{11}_{\Lambda\Lambda}$Be → ${}^{5}_{\Lambda}$He + $d$ + $d$ + 2$n$ | $< 121.01 \pm 0.42$ |
| ${}^{11}_{\Lambda\Lambda}$Be → ${}^{5}_{\Lambda}$He + $t$ + $p$ + 2$n$ | $< 125.47 \pm 0.45$ |
| ${}^{13}_{\Lambda\Lambda}$B → ${}^{5}_{\Lambda}$He + $p$ + ${}^{4}$He + 3$n$ | $< 116.24 \pm 0.38$ |
| ${}^{13}_{\Lambda\Lambda}$B → ${}^{5}_{\Lambda}$He + $d$ + ${}^{3}$He + 3$n$ | $< 99.80 \pm 0.39$ |
| ${}^{13}_{\Lambda\Lambda}$B → ${}^{5}_{\Lambda}$He + $d$ + ${}^{4}$He + 2$n$ | $< 112.57 \pm 0.63$ |
| ${}^{13}_{\Lambda\Lambda}$B → ${}^{5}_{\Lambda}$He + $t$ + ${}^{3}$He + 2$n$ | $< 100.98 \pm 0.64$ |
| ${}^{13}_{\Lambda\Lambda}$B → ${}^{5}_{\Lambda}$He + ${}^{4}$He + $p$ + 3$n$ | $< 119.33 \pm 0.51$ |
| ${}^{13}_{\Lambda\Lambda}$B → ${}^{5}_{\Lambda}$He + ${}^{3}$He + $d$ + 3$n$ | $< 101.26 \pm 0.46$ |
| ${}^{13}_{\Lambda\Lambda}$B → ${}^{5}_{\Lambda}$He + ${}^{4}$He + $d$ + 2$n$ | $< 114.45 \pm 0.78$ |
| ${}^{13}_{\Lambda\Lambda}$B → ${}^{5}_{\Lambda}$He + ${}^{3}$He + $t$ + 2$n$ | $< 101.59 \pm 0.71$ |
| ${}^{14}_{\Lambda\Lambda}$C → ${}^{5}_{\Lambda}$He + $p$ + ${}^{6}$Li + 2$n$ | $< 76.35 \pm 0.88$ |
| ${}^{14}_{\Lambda\Lambda}$C → ${}^{5}_{\Lambda}$He + ${}^{3}$He + ${}^{3}$He + 3$n$ | $< 84.27 \pm 0.58$ |
| ${}^{14}_{\Lambda\Lambda}$C → ${}^{5}_{\Lambda}$He + ${}^{3}$He + ${}^{4}$He + 2$n$ | $< 96.65 \pm 0.94$ |
| ${}^{14}_{\Lambda\Lambda}$C → ${}^{5}_{\Lambda}$He + ${}^{4}$He + ${}^{3}$He + 2$n$ | $< 97.39 \pm 1.00$ |
| ${}^{14}_{\Lambda\Lambda}$C → ${}^{5}_{\Lambda}$He + ${}^{6}$Li + $p$ + 2$n$ | $< 81.37 \pm 1.18$ |

Only decay modes of ${}^{11}_{\Lambda\Lambda}$Be, ${}^{13}_{\Lambda\Lambda}$B and ${}^{14}_{\Lambda\Lambda}$C are listed. The result of kinematic fitting decides errors in $\Delta B_{\Lambda\Lambda}$ values.

while the Demachiyanagi event from E373 yielded $B_{\Lambda\Lambda} = 27.81^{+3.16}_{-2.02}$ MeV and $\Delta B_{\Lambda\Lambda} = 5.07^{+3.17}_{-2.03}$ MeV[20]. Although the divergences of those $\Delta B_{\Lambda\Lambda}$ values are within 3$\sigma$ from the well-established $\Delta B_{\Lambda\Lambda} = 0.67 \pm 0.17$ MeV for the uniquely identified ${}^{6}_{\Lambda\Lambda}$He[20], they suggests a possible nuclear species dependence of $\Delta B_{\Lambda\Lambda}$. However, the ambiguity surrounding the nuclide and the mass number of the previously observed double-Λ hypernuclei precluded a firm conclusion. This study provides the first unambiguous measurement of $B_{\Lambda\Lambda}$ and $\Delta B_{\Lambda\Lambda}$ for ${}^{13}_{\Lambda\Lambda}$B, thereby eliminating nuclide identification uncertainties. Notably, our $\Delta B_{\Lambda\Lambda}$ value was larger than that of ${}^{6}_{\Lambda\Lambda}$He, providing the first direct indication that the strength of the ΛΛ interaction depends on the nuclear medium.

At present, the machine learning model focuses primarily on the topological features of the ${}^{6}_{\Lambda\Lambda}$He event in nuclear emulsion. Because the training dataset is based on simulated events, potential biases may arise from differences between the simulated and real emulsion data, including variations in background structures. The general topology of other possible double-strangeness events will need to be incorporated in future training to improve model robustness against unforeseen backgrounds. As more double-strangeness candidates are discovered and confirmed through conventional visual and kinematic analyses, the model can be progressively validated and refined. Extending the machine learning framework to include kinematic information will be an important step toward establishing a more comprehensive and reliable identification method.

An AI-driven double-strangeness factory will transform our understanding of baryonic interactions in multi-strangeness systems. Using this approach, a large-scale analysis of nuclear emulsion will reveal a vast population of double-strangeness hypernuclei, enabling high-precision studies of ΛΛ interactions, quantum three-body forces through ΛΛ-Ξ$N$ coupling, and exotic multi-baryon states. These breakthroughs will provide deeper insights into the composition of neutron star cores and the possible existence of the H-dibaryon, marking a significant step forward in double-strangeness hypernuclear physics.

# Methods

In this study[37], we employed GANs[38] and Geant4 Monte Carlo simulations[39] to generate training data for the Mask R-CNN[40], which was used to detect double-Λ hypernuclear events in nuclear emulsion. Mask R-CNN training requires images with objects of double-Λ hypernuclear events and their corresponding mask images. However, for double-Λ hypernuclear events, there are insufficient data to train the model, as only one event has been uniquely identified to date. To address this issue, Geant4 simulations were used to generate double-Λ hypernuclear events in nuclear emulsion, as illustrated in Fig. 3a. Combined with background tracks, these simulated events were processed using an image-style transformation via pix2pix[38] with GANs to produce training images. Mask images were automatically generated from Geant4 track information. The Mask-R CNN model was subsequently trained using the generated training datasets, including images containing double-Λ hypernuclear objects and mask images, as illustrated in Fig. 3b, c, respectively. The model evaluation detailed in Section "Model performance and event detection with AI" demonstrated the effective detection of double-Λ hypernuclear events in the produced images. Notably, the model accurately detected and segmented the NAGARA event with a confidence score of 0.974, as shown in Fig. 3f. Subsequently, we applied the trained model to 0.2% of the E07 emulsion data, which led to the discovery and unique identification of a double-Λ hypernuclear event, as discussed in Section "Result".

## Training data prepared with Geant4 simulation and generative AI

In Geant4 Monte Carlo simulations, the composition of the nuclear emulsion was replicated by referring to the emulsion layer used in the J-PARC E07 experiment. For the double-Λ hypernuclear events generated, we first considered the case of ${}^{6}_{\Lambda\Lambda}$He and its sequential decay. As illustrated in Fig. 3a, ${}^{6}_{\Lambda\Lambda}$He is produced by the $\Xi^-$ capture of ${}^{12}$C in the nuclear emulsion at vertex A. We assumed that $\Xi^-$ is bound in the 3D atomic orbit of ${}^{12}$C with a binding energy of 0.13 MeV[50]. As shown in Fig. 3b, the thickness of each track in the nuclear emulsion, which is related to the grain density, was calculated and displayed for different tracks based on their velocity and angle. Because the tracks in the nuclear emulsion were recorded with three-dimensional information, the trajectories were converted into three different colors, as discussed in ref. 37. To ensure an accurate classification and detection performance, negative samples, $K^-$ beam interaction events, were generated using the JAM package[54] as main background events. Additional background tracks were extracted from the microscopic images of the E07 emulsion data using an image filter and binarization[37].

After generating all the particle tracks in the emulsion images, image-style transfer using GANs was applied to generate emulsion images that closely mimicked real emulsion images. Based on the capabilities of GANs, the pix2pix model was employed to convert the RGB image into an image similar to a real emulsion image, as shown in Fig. 3b. The pix2pix model is an implementation of the conditional GANs framework, which shows a significant improvement in performance for image-style transformation, especially for high-resolution images. The parameters used for training the pix2pix model in this study are aligned with those specified in our previous study[55]. The image produced by the trained pix2pix model in Fig. 3b, combined with the corresponding mask images in Fig. 3c, served as training data for the object detection model, Mask R-CNN.

## Model performance and event detection with AI

After training with the produced data, the developed model can detect double-Λ hypernuclear events in both the produced and actual nuclear emulsion images. For the produced images, the method achieved detection efficiencies of 93.8% and 82.0% for ${}^{6}_{\Lambda\Lambda}$He and ${}^{5}_{\Lambda\Lambda}$H, respectively, with corresponding purities of 98.2–98.3%. Furthermore, it successfully detected the NAGARA event with a confidence score of 0.974. When applied to E07 emulsion images, the method drastically reduced the background images to 0.17% of the original level. It successfully detected six candidates of double-Λ hypernuclear events in over 0.2% of the entire nuclear emulsion dataset from the E07 experiment. The number of detected candidates suggests that more than 2000 double-strangeness hypernuclear events were recorded in the entire dataset. The proposed method demonstrates great potential for use across the entire E07 nuclear emulsion dataset, potentially improving the visual inspection efficiency by approximately 500 times. Details of the developed method are summarized in ref. 37.

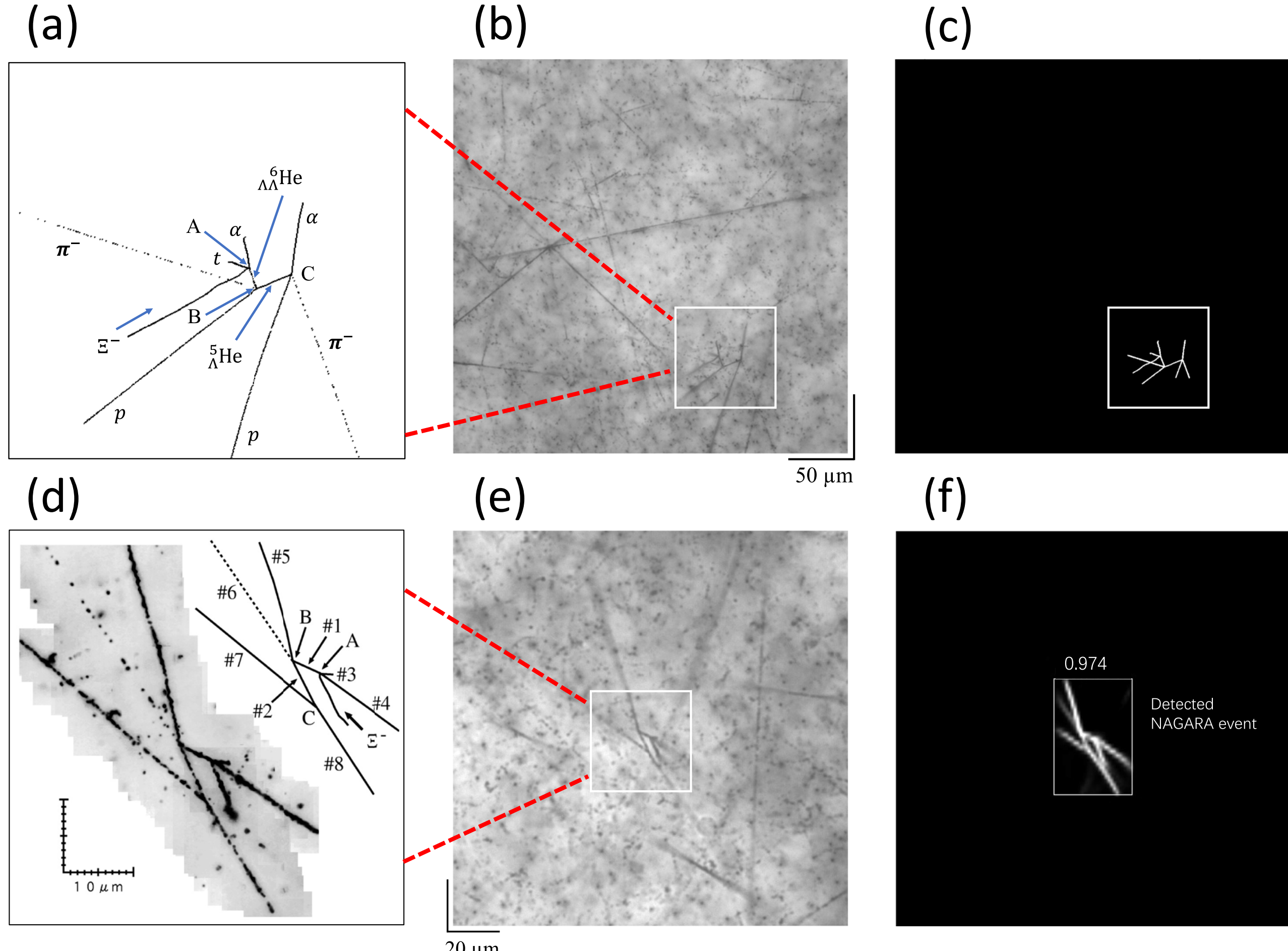

**Fig. 3 | Training image and model performance on NAGARA event. a** displays the topology of a double-$\Lambda$ hypernuclear event generated with Geant4 Monte Carlo simulations. **O**ne of the example images for the training of the machine learning model (Mask R-CNN), including an image of a goal object (**b**) and the images with its mask image (**c**). **d** shows the images of the NAGARA event taken from ref. 19, which is the only uniquely identified double-$\Lambda$ hypernuclear event thus far. **e** is the input image of the Nagara event for the developed model. **f** shows the detection result of the developed model for the NAGARA event.

### Calibration of range–energy relationship

Before the kinematic analysis of the event, the range-energy relationship in the emulsion sheet was calibrated using $\alpha$ tracks, which have a monochromatic energy of 8.784 MeV from the decay of $^{212}$Po existing in the emulsion. The $\alpha$ track can be identified in the thorium series isotopes because it has the largest kinetic energy. Such $\alpha$ decay chains were searched for around the observed events using the so-called overall scanning method[56]. In total, 217 $\alpha$ tracks were scanned in the emulsion sheet for calibration. The relationship between the ranges and the kinetic energies of the charged particles was obtained using the range-energy formula given by Barkas et al.[57,58]. The equivalent density of the emulsion sheet was determined to be 3.622 ± 0.014 g/cm$^3$ for the emulsion layer in which the newly observed event was detected. The ranges of particles needed to be corrected for the shrinkage effect because the emulsion layers shrunk perpendicular to the surface due to photographic development. The range of 217 $\alpha$ tracks can also determine this shrinkage factor. Finally, the mean range of $\alpha$ tracks was obtained as 49.57 ± 0.12 μm and the shrinkage factor was corrected to be 1.93 ± 0.01.

The kinetic energy of the $\pi^-$ particle was determined using a range–energy calibration based on $\mu^+$ particles from $\pi^+$ meson decays at rest, which provided a monochromatic energy of 4.1205 MeV. Although ATIMA[59] calculations indicated a difference in composition between E07 and standard emulsions, primarily affecting long-range particles such as $\pi^-$, the density was determined to be 3.379 ± 0.006 g/cm$^3$ using 160 $\mu^+$ tracks and ATIMA calculations[60]. This density value was then used to calculate the $\pi^-$ particle's kinetic energy.

## Data availability

The data used for training Mask R-CNN model generated in this study have been deposited in the Github repository under accession code https://github.com/ayumisalt/mask_rcnn_emulsion.git[61].

## Acknowledgements

This work was supported by JSPS KAKENHI Grant Numbers JP25H00404, JP16H02180, JP20H00155, JP18H05403, JP19H05147 (Grant-in-Aid for Scientific Research on Innovative Areas 6005), JP25H01550 (Grant-in-Aid for Transformative Research Areas), JP25K17415 (Grant-in-Aid for Early-Career Scientists), and JP23K19051 (Grant-in-Aid for Research Activity Start-up). The authors acknowledge support from Proyectos I+D+i 2020 (ref: PID2020-118009GA-I00); Proyectos I+D+i 2022 (ref: PID2022-140162NB-I00); and Proyecto Consolidación Investigadora 2022 (ref: CNS2022-135768); as well as from Grants 2019-T1/TIC-13194 and 2023-5A/TIC-28925 under the Atracción de Talento Investigador program of the Community of Madrid. H.W. acknowledges support from the Major Science and Technology Projects in Gansu Province under Grant No. 24ZD13GA005. The authors thank the J-PARC E07 collaboration for providing the emulsion sheets. The authors thank Michi Ando, Risa Kobayashi, and Yuki Mochizuki of RIKEN and Yoko Tsuchii of Gifu University for their technical support during the mining events in the J-PARC E07 nuclear emulsions. The authors thank Yukiko Kurakata of RIKEN for the administrative work.

## Author contributions

Y.H. and T.R.S. implemented the original draft. Y.H., H.E., A.K., E.L., C.R., M.T., and J.Y. developed the analysis method, including machine learning. Y.H., T.R.S., H.E., A.K., K.N., C.R., Y.K.T., and H.W. performed kinematics and physics analyses. Y.H., H.E., and A.K. visually inspected the event candidates using microscopes. Y.H., T.R.S., H.E., A.K., Y.G., E.L., K.N., C.R., M.T., Y.K.T., H.W., A.Y., J.Y. and H.Z. discussed the results and reviewed the manuscript. T.R.S. coordinated and supervised this project.

## Competing interests

The authors declare no competing interests.

## Additional information

**Supplementary information** The online version contains supplementary material available at https://doi.org/10.1038/s41467-025-66517-x.

**Correspondence** and requests for materials should be addressed to Yan He or Takehiko R. Saito.

**Peer review information** *Nature Communications* thanks Tom Reichert and the other anonymous reviewer(s) for their contribution to the peer review of this work. A peer review file is available.